\documentclass[prl,aps,amssymb,amsmath,reprint,superscriptaddress,showpacs,groupedaddress]{revtex4-1}
\usepackage{bm}
\usepackage{graphicx}
\usepackage{color}

\usepackage{amssymb}
\usepackage{hyperref}

\begin{document}

\title{Interplay of Kondo effect and strong spin-orbit coupling in multi-hole 
ultraclean carbon nanotubes}

\author{J. P. Cleuziou}
\author{N. V. N'Guyen}
\author{S. Florens}
\author{W. Wernsdorfer}
\affiliation{Institut N\'eel, CNRS et Universit\'e Joseph Fourier BP 166, 38042
Grenoble Cedex 9, France}

\begin{abstract}
We report on cotunneling spectroscopy magnetoconductance measurements of
multi-hole ultraclean carbon nanotube quantum dots in the SU(4) Kondo regime
with strong spin-orbit coupling. Successive shells show a gradual weakening of the
Kondo effect with respect to the spin-orbital splittings, leading to an
evolution from SU(4) to SU(2) symmetry with a suppressed conductance at half
shell filling. The extracted energy level spectrum, overally consistent with
negligible disorder in the nanotube, shows in the half filled case large
renormalizations due to Coulombian effects.
\end{abstract}

\maketitle

The interplay between electron orbital and spin degrees of freedom in
nanostructures is the current subject of intense studies. In this context,
carbon nanotubes (CNTs) have attracted a considerable interest due to quite
unique electronic properties \cite{hamada1} making them attractive candidates
for spintronics devices \cite{sahoo1, urdampilleta1} and spin qubits
\cite{churchill1,pei1}. In particular, single-electron tunneling spectroscopy in
ultraclean CNT devices \cite{kuemmeth1,kuemmeth2} has shown the important role
of the curvature induced spin-orbit interaction (SOI) \cite{ando1,huertas1},
which partially splits the fourfold degenerate electron shell level structure by
an effective energy gap $\Delta_{SO}$. This result was confirmed in the many
electron regime, including a significant disorder induced orbital mixing
$\Delta_{KK'}\gg \Delta_{SO}$ \cite{jespersen1}.

On the other hand, it is well established that quantum dots with sufficiently
strong tunnel couplings to the metallic leads exhibit rich Kondo physics
\cite{florens1}. In particular, for a CNT having the fourfold level degeneracy,
the dot spin and orbital degrees of freedom become strongly coupled to those of
the leads (assuming conserved orbital quantum numbers during the tunneling
processes), thereby forming a highly entangled SU(4) Kondo ground state
\cite{choi1,anders1} with an enhanced Kondo temperature $T_{K}$
\cite{jarillo1,makarovski1}. By taking into account the SOI in the CNT energy
spectrum, a mutual interplay between the SOI and Kondo correlations is expected
\cite{galpin1,fang1}, resulting in a crossover from SU(4)
\cite{makarovski1,makarovski2} ($k_{B}T_{K} \gg \Delta_{SO}$) to more
conventional SU(2) Kondo effects \cite{nygard1} ($k_{B}T_{K} \ll \Delta_{SO}$)
at zero magnetic field. While some deviations of SU(4) Kondo physics
related to the SOI have been discussed theoretically \cite{galpin1} in connexion
to previous experiments \cite{makarovski1, jarillo1}, the full detailed experimental
study of disorder-free CNT quantum dots in the Kondo regime with 
well-resolved SOI split energy spectrum is still missing.

\begin{figure}[H!b]
\includegraphics[scale=0.86]{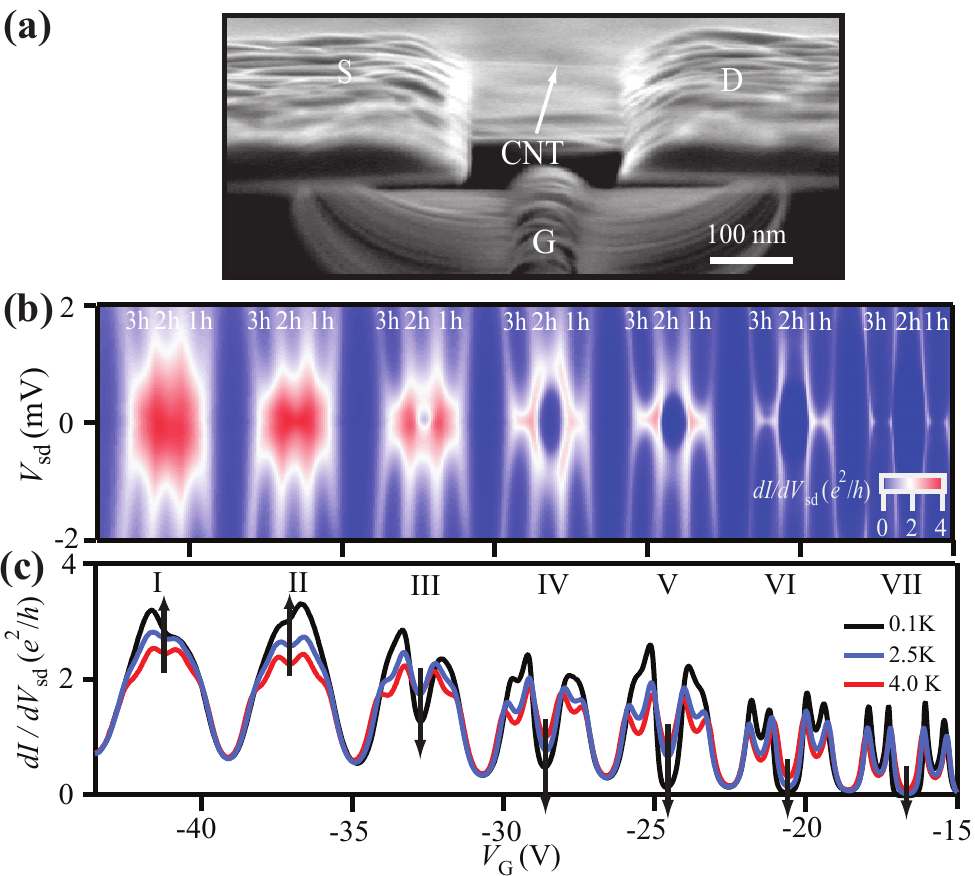}
\centering 
\caption{(color online). (a) Scanning electron microscopy of an ultraclean
nanotube device similar to the one used in the measurements. (b) Differential
conductance map $dI/dV_{sd}$ versus gate voltage $V_{G}$ and bias voltage
$V_{sd}$ at $T$ $\approx$ 30 mK. (c) Linear differential conductance versus $V_{G}$, measured at different temperatures.}
\label{fig:Fig1New}
\end{figure}

In this Letter, we study by means of cotunneling magneto-spectroscopy under a
parallel magnetic field $B_{\parallel}$ (see Ref.~\cite{SupInfo}, section 1.2 for the actual
alignment relative to the CNT axis in Fig.~\ref{fig:Fig1New}(a)) the orbital structure
and the Kondo effects in an ultraclean CNT quantum dot, with an energy level
structure dominated by the SOI ($\Delta_{SO}\gg \Delta_{KK'}$). This complements
recent studies performed in the sequential tunneling regime where Kondo physics
was absent \cite{kuemmeth1,kuemmeth2}, while previous cotunneling measurements
were made on disorder-dominated samples~\cite{jespersen1,grove1}. One striking finding is the strong
renormalization of the effective SOI splitting at half shell filling, denoted
$\Delta_{SO}^\star$ (see below), resulting systematically in very different SOI
energy scales for odd and even hole numbers ($\Delta_{SO}^\star > \Delta_{SO}$).
In the weak Kondo regime ($k_{B}T_{K}$ $\ll$ $\Delta_{SO}^\star$), the Kondo
effect is quenched for an even charge on the dot, as predicted
theoretically~\cite{galpin1}. As the dot tunnel coupling is then increased with
larger gate voltages, the Kondo effect is greatly enhanced and Kondo
correlations tend to restore the full SU(4) Kondo symmetry
\cite{jarillo1,makarovski1}. For intermediate gate voltages, we are thus able to
study the regime of comparable Kondo and orbital energy scales ($k_{B}T_{K}
\simeq \Delta_{SO}^\star$) where the orbital structure competes with SU(4) Kondo
physics~\cite{galpin1}.

The quantum dot studied here is based on an ultraclean CNT \cite{kuemmeth1}
grown on top of predefined source (S) and drain (D) Pt metal contacts and
suspended over a gate electrode~\cite{cao1} (Fig.~\ref{fig:Fig1New}(a)). A
different device, showing similar physics, is further discussed in
Ref.~\cite{SupInfo}, section 2. Applying a voltage on the gate shifts the chemical
potential of the single CNT quantum dot ($\approx$ 200-nm-long)
\cite{cleuziou2}, as depicted in the device stability diagram of
Fig.~\ref{fig:Fig1New}(b). As negative $V_{G}$ shifts the Fermi level below the
edge of the nanotube bandgap ($E_{g} \approx$ 300 meV), additional holes
(typically from 10 to 40 here) are subsequently added to the CNT valence band and the
conductance displays a series of Coulomb peaks for each additional hole,
connected by either Coulomb blockade valleys or Kondo ridges at zero bias
\cite{nygard1}. The four hole periodicity in the charge stability diagram of
Fig.~\ref{fig:Fig1New}(b) demonstrates that the energy level spectrum consists of
four-fold nearly degenerate shells separated by the shell mean level spacing
$\Delta E_\mathrm{shell}$ $\approx$ 4 meV (see \cite{SupInfo}, section 1.1), approximately
constant for the shells I-VII depicted in Fig.~\ref{fig:Fig1New}(b).

As $V_{G}$ is raised towards more negative values, the contact p-n junctions
become more transparent \cite{makarovski2}, resulting in a gradual enhancement
of the conductance close to the saturation limit. For the highest tunnel
couplings, reached for the shells I and II, the conductance is greatly enhanced
for all partial shell fillings due to the SU(4) Kondo effect, resulting in Kondo
ridges in the middle of the $N$ = $1h$, $2h$ and $3h$ Coulomb diamonds
\cite{makarovski2}. However, as $T_{K}$ decreases for the next shells, the
zero-bias conductance at half filling is gradually quenched in such a way that
zero bias Kondo ridges are observed for odd numbers of holes only
(Fig.~\ref{fig:Fig1New}(b)). This behavior is confirmed by the $T$-dependence of
the zero bias conductance with gate voltage, depicted in Fig.~\ref{fig:Fig1New}(c),
showing a decreasing conductance with lowering temperature at half filling of
the shells III-VII, in opposite behavior to the shells I-II. 
As we will demonstrate by cotunneling level spectroscopy, the disapperance of
the SU(4) Kondo effect is caused by significant splittings of the even hole
energy spectrum, in the absence of disorder induced orbital mixing, preventing spin-orbital level degeneracies to form for even
number of holes. 

\begin{figure}[!Ht]
\includegraphics[scale=0.5]{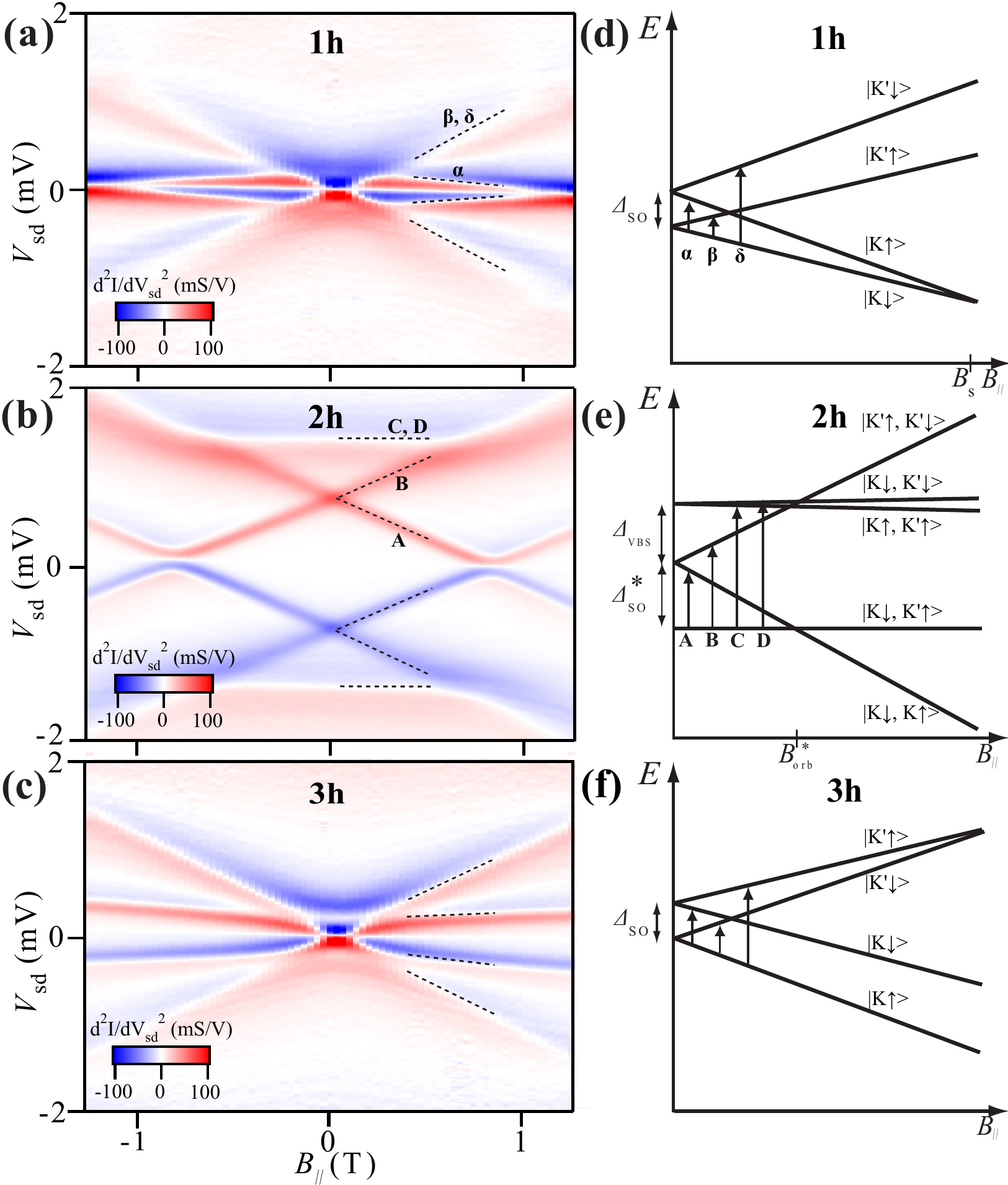}
\caption{(color online). (a-c) Current second derivative $d^{2}I/dV_{sd}^{2}$
versus bias voltage $V_{sd}$ and axial magnetic field $B_{\parallel}$ in the
middle of the $N$ = $1h$, $2h$, and $3h$ charge states of the shell VI.
(d-f) Schematic representation of the expected low energy spin-orbital levels as
a function of $B_{\parallel}$, corresponding to (a-c).}
\label{fig:Fig2New}
\end{figure}

We first discuss the level spectrum and the evolution of the
Kondo resonance for $N$ = $1h$ under parallel magnetic field $B_{\parallel}$ in
shell VI, corresponding to about 20 holes trapped in the valence band
(Fig.~\ref{fig:Fig2New}(a)). The Kondo peak mainly splits according to two most
visible peaks, highlighted by the dashed black lines in Fig.~\ref{fig:Fig2New}(a),
evolving in $B_{\parallel}$ with significantly different
$dV_{sd}/dB_{\parallel}$ slopes, in good agreement with an energy level spectrum
including the curvature induced SOI (Fig.~\ref{fig:Fig2New}(d)). 
The $N$ = $1h$ states at $B_{\parallel}$ = 0 consist in a pair of Kramers
doublets formed by spin-orbital states with an antiparallel alignment of the
orbital and spin magnetic moments~\cite{kuemmeth1} [namely
$\left\{|K\downarrow\rangle, |K'\uparrow\rangle\right\}$ for the lowest doublet,
and $\left\{|K\uparrow\rangle,|K'\downarrow\rangle\right\}$ for the highest
one], that is split by the effective spin-orbit gap $\Delta_{SO}$. Turning on
the $B_{\parallel}$-field separates the two energy doublets into four
spin-orbital states with slopes proportional to either $g_{orb}\pm g_{s}$ or
$-g_{orb}\pm g_{s}$, with $g_{orb}\gg g_{s}$ the orbital and spin $g$-factors
respectively~\cite{minot1}. This one-hole level scheme results in three
$B_{\parallel}$-field dependent excitations~\cite{jespersen1}, denoted $\alpha$,
$\beta$ and $\delta$ in Fig.~\ref{fig:Fig2New}(d). The transitions are better
resolved by plotting in Figs.~\ref{fig:Fig2New}(a-c) the second derivative of the
current, $d^2I/d^2V_{sd}$, noting that the excitations correspond to lines of
either extrema or zeros, depending on the degree of renormalization of the
cotunneling processes at play~\cite{paaske1}. Extrapolating the conductance peak
associated to $\alpha$ at $B_{\parallel}$ = 0 in Fig.~\ref{fig:Fig2New}(a) gives
the amplitude of the effective spin-orbit splitting $\Delta_{SO} \approx$ 0.24
meV for $N$ = $1h$, which is of the same order of magnitude than previous
findings~\cite{kuemmeth1, kuemmeth2, jespersen1}. The energy dependence of the
conductance peak associated to $\alpha$ also results in a SU(2) Kondo effect at
the finite parallel magnetic field $B_{s}=\Delta_{SO}/g_{s}\mu_{B}$ due to the
$B_{\parallel}$-field induced level crossing for $N$ = $1h$. The other main
visible conductance peak in Fig.~\ref{fig:Fig2New}(a) takes into account the two
other expected $\beta$ and $\delta$ excitations (Fig.~\ref{fig:Fig2New}(d)), which
are difficult to separate spectroscopically because of the tunnel-induced
broadening of the two associated conductance peaks~\cite{jarillo1}. The large
and positive slope of $dV_{sd}/dB_{\parallel}$ for $\beta$ and $\delta$, roughly
proportional to $g_{orb}$, reveals the dominant contribution from the orbital
$g$-factor. In contrast, the dependence in $g_{orb}$ drops for the $\alpha$
transition, leading to a weak and negatively dispersing transition line in the
conductance due to the spin Zeeman effect only, as indeed expected from the
level spectrum of Fig.~\ref{fig:Fig2New}(d).

The $N$ = $3h$ theoretical energy level spectrum (Fig.~\ref{fig:Fig2New}(f)) is
expected to be equivalent to the $N$ = $1h$ one (Fig.~\ref{fig:Fig2New}(d)) by just
exchanging the two doublets at $B_{\parallel}$ = 0~\cite{jespersen1}
(particle-hole symmetry), and indeed the measurement of Fig.~\ref{fig:Fig2New}(c)
is in good agreement with the level scheme of Fig.~\ref{fig:Fig2New}(f). We find
here a spin-orbit gap $\Delta_{SO} \approx$ 0.23 meV, very close to the value
measured for $N$ = $1h$ ($\Delta_{SO} \approx$ 0.24 meV). Contrary to the
situation with a single hole, we note the absence of $B_{\parallel}$-field
induced crossing at low-energy due to the different $N=3h$ level scheme~\cite{makarovski1,jespersen1,galpin1}.

After considering the SOI level structure for $N$ = $1h$ and $3h$, we now examine the situation at half shell filling ($N$ = $2h$), depicted in
Figs.~\ref{fig:Fig2New}(b,e). The $N$ = $2h$ energy states can be readily built from linear combinations of the single hole states, leading to six independent configurations. At $B_{\parallel}$ = 0, the singlet-like ground state (denoted $\left|K\downarrow,K'\uparrow\right\rangle$ in Fig.~\ref{fig:Fig2New}(e)) corresponds to the two holes occupying the low-energy Kramers doublet, and sets the origin of the cotunneling excitations~\cite{jespersen1}. The four excitations, denoted A, B, C and D in Figs.~\ref{fig:Fig2New}(b,e) refers to the transitions from the ground state to the first four excited states where only one hole lies in the high-energy Kramers doublet. The expected sixth state ($\left|K\uparrow, K' \downarrow\right\rangle$) is not visible experimentally in Fig.~\ref{fig:Fig2New}(b)~\cite{jespersen1} since it would imply simultaneous spin
flips of both holes. The excitations (A, B) appear as $dI/dV_{sd}$ conductance
steps (peaks in $d^{2}I/dV_{sd}^{2}$ of Fig.~\ref{fig:Fig2New}(b)), with
$\left|dV_{sd}/dB_{\parallel}\right|$ average slopes given by the orbital Zeeman
effect. We deduce here an orbital $g$-factor $g_{orb}$ $\approx$ 15.5$\pm$0.5
(equivalently an orbital magnetic moment $\mu_{orb}$ = $g_{orb}\mu_{B}/2$ = 0.45
meV.T$^{-1}$), in agreement with the $B_{\parallel}$-field orbital splittings of
$N$ = $1h$ and $3h$ (Figs.~\ref{fig:Fig2New}(a,c)). The excitations (C, D) merge
together into an enhanced $dI/dV_{sd}$ Kondo peak at finite bias~\cite{paaske1},
such as both excitations cannot be separately distinguished under the
$B_{\parallel}$-field. At $B_{\parallel}$ = $B^{\ast}_{orb}$ =
$\Delta^{\ast}_{SO}$/$g_{orb}\mu_{B}$ (see below for the definition of
$\Delta^{\ast}_{SO}$), a singlet/triplet-like level crossing
occurs~\cite{fang1}, leading to a new ground state configuration for $B_{\parallel} > B^{\ast}_{orb}$. From the avoided level crossing between the $\left|K\downarrow,K\uparrow\right\rangle$ and $\left|K\downarrow,K'\uparrow\right\rangle$ states at $B^{\ast}_{orb}$ in
Fig.~\ref{fig:Fig2New}(b), we estimate a very weak disorder-induced valley mixing ($\Delta_{KK'} \approx$ 0.04 meV), in contrast to previous
cotunneling studies~\cite{jespersen1,grove1}, where disorder was largely dominating the SOI. Our measurements therefore complement sequential tunneling spectroscopies in the ultraclean regime~\cite{kuemmeth2}.

One striking observation made in our device is the huge effective SOI
splitting for $N$ = $2h$, $\Delta^{\ast}_{SO}\approx$ 0.75 meV in
Figs.~\ref{fig:Fig2New}(b,e), as defined by the energy of the first excitations
at $B_{\parallel}$ = 0. This strong value cannot be accounted for the curvature
induced SOI only ($\Delta_{SO}$ $\approx$ $\Delta^{\ast}_{SO}$/3) since the
latter is mainly related to the specific CNT bandstructure and thus evolves
continuously from odd to even valleys~\cite{jespersen1}. Here, the enhancement
of $\Delta^{\ast}_{SO}$ with respect to $\Delta_{SO}$ probably originates from
short range Coulomb interactions, well visible at half shell filling. More
specifically, recent theoretical studies~\cite{rontani1,wunsch1,rontani3}
underlined the important role of Coulomb inter-valley backscattering (VBS)
processes, resulting in an additional lifting of degeneracies in the spin-orbit level
structure. This interpretation is indeed confirmed by the significant effective
splitting $\Delta_{VBS}$ $\approx$ 0.67 meV of the excited
quadruplet~\cite{rontani1,rontani3}, associated to the transitions (A, B) and
(C, D), regrouped into two energy doublets at $B_{\parallel}$ = 0,
depending whether the two holes have equal or opposite isospins. While the sign of
this interaction is in good agreement with previous observation~\cite{kuemmeth2}, the large magnitude of the effective $\Delta_{VBS}$ is dramatically larger in our experiment ($\Delta_{VBS}$ $\approx$ 0.19 meV in Ref.~\cite{kuemmeth2}).
Our observations put strong challenges for current theoretical
works~\cite{rontani1,wunsch1,rontani3}, and suggest possible neglect of 
many-body effects related to screening processes, since the quantitative determination of exchange constants
is a notoriously difficult problem~\cite{handrick,lepetit}.

\begin{figure}[!Ht]
\includegraphics[scale=0.73]{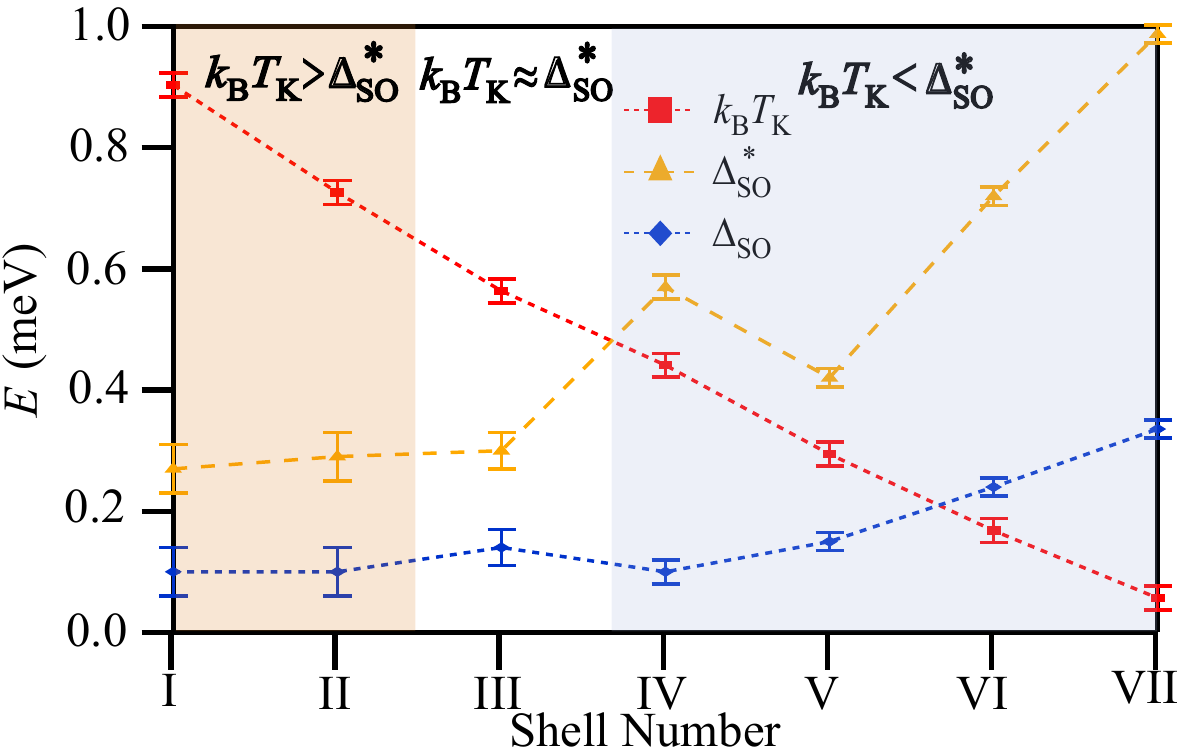}
\caption{(color online). Evolution of Kondo temperature $T_{K}$, one-hole spin-orbit gap $\Delta_{SO}$ and two-hole effective gap $\Delta^{\ast}_{SO}$ for the shells I-VII.}
\label{fig:Fig3New}
\end{figure}

After identifying the influence of the SOI in the energy level structure of a given hole shell (Fig.~\ref{fig:Fig2New}), we now show its consequence on the Kondo effect seen in Fig.~\ref{fig:Fig1New}. Figure~\ref{fig:Fig3New} depicts the amplitudes of $\Delta_{SO}$ and $\Delta^{\ast}_{SO}$, measured respectively for the $N$ = $1h$ and $2h$ partial shell fillings, in the seven available shells of Fig.~\ref{fig:Fig1New}(b). The magnitude of the SOI gap $\Delta_{SO}$ increases with decreasing gate voltage from the shells I to VII, as expected for curvature
induced SOI in the valence band~\cite{jespersen1,jespersen2}. Interestingly, $\Delta_{SO}$ and $\Delta^{\ast}_{SO}$ follow a comparable gate dependence, with a ratio $\Delta^{\ast}_{SO}$/$\Delta_{SO}$ $\approx$ 3 that is approximately constant for all shells considered in Fig.~\ref{fig:Fig3New} (except for shell IV). For the successive hole shells, the effective SOI gap $\Delta^{\ast}_{SO}$ is compared  with the Kondo temperature $T_{K}$, estimated from the half width at half maximum of the $dI/dV_{sd}$ versus $V_{sd}$ curves, recorded in the middle
of the $N$ = $1h$ diamond (see Fig.~\ref{fig:Fig3New}). We find that the Kondo temperature gradually decreases from the shells I to VII, in agreement with the evolution of the $U$/$\Gamma$ ratios with the gate voltage. The opposite evolutions of the Kondo temperature and the SOI with $V_{G}$ (Fig.~\ref{fig:Fig3New}) explain now the different transport regimes previously seen in Fig.~\ref{fig:Fig1New}, depending whether Kondo or spin-orbit physics prevail, as we discuss now.

\begin{figure}
\includegraphics[scale=0.87]{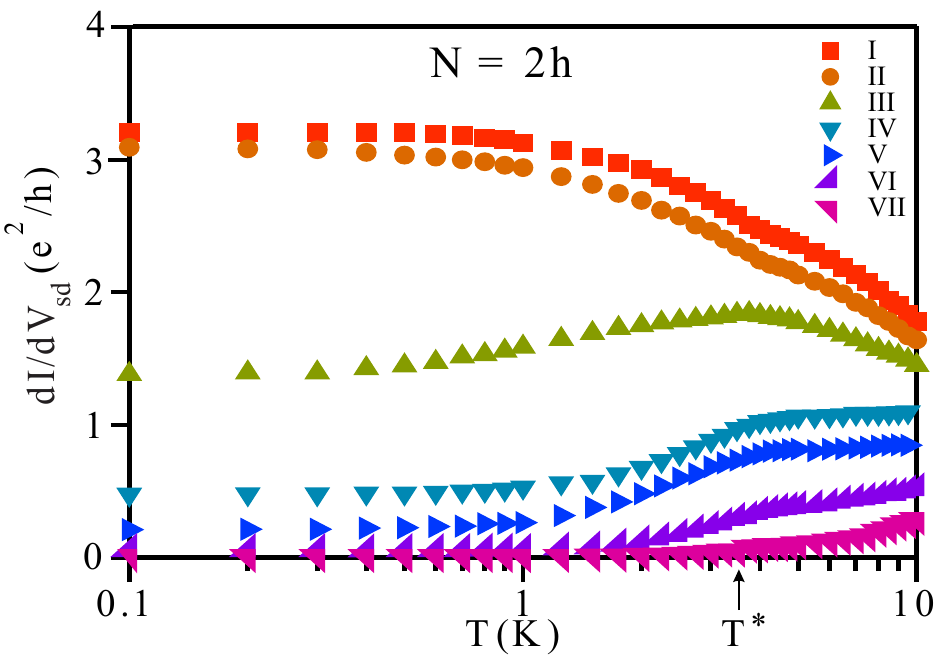}
\caption{(color online). Zero bias differential conductance $dI/dV_{sd}$ versus
$T$ in the center of even hole diamonds ($N$ = $2h$) in the successive shells
(I-VII), from $T$ = 100 mK up to 10 K.}
\label{fig:Fig4New}
\end{figure}

The competition between Kondo and SOI in the two-hole sector is better seen by the full dependence of the conductance versus temperature for all shells in Fig.~\ref{fig:Fig4New}. For the first two $N$ = $2h$ shells, the conductance saturates at $T < T_{K}$ up to 3.1-3.2 $e^{2}/h$, close but below the unitary limit 4 $e^{2}/h$. Due to the high conductance for odd valleys ($dI/dV_{sd}$ $\approx$ 2$e^2/h$), see Fig.~\ref{fig:Fig1New}c, we can rule out source-drain asymmetries of the tunnel barriers to explain this reduction of conductance. In addition, orbital asymmetries of the tunnel couplings~\cite{kirsanskas1,grove1} also are unlikely here, due to the clearly-evidenced weak disorder in  our sample. Thus, we can attribute the weakly broken SU(4) Kondo state in shells I-II to the small but sizeable (in comparison to $T_K$) effective spin-orbit coupling  $\Delta^{\ast}_{SO}$, as seen in Fig.~\ref{fig:Fig3New}. This interpretation is confirmed by the next two-hole shell III, where the renormalized SOI and Kondo energies become comparable and compete, resulting in a marked drop of conductance down to 1.4 $e^2/h$ at low temperature. This rapid but not total quenching of the SU(4) Kondo effect is in qualitative agreement with theoretical expectations~\cite{galpin1}, and results for the shell III in a conductance maximum ($dI/dV_{sd}$ $\approx$ 1.9 $e^2/h$) at the temperature $T^{\ast}\approx \Delta^{\ast}_{SO}/k_B$. By further weakening the Kondo temperature in shells IV-VII, one observes a complete suppression of the Kondo effect (cotunneling regime), with a purely monotonous $T$-dependence of the conductance, due to the predominence of orbital splitting which totally remove the degeneracies in the system. Our ultraclean CNT with sizeable tunnel couplings to the metal leads thus offers a complete span from SU(4) to SU(2) behavior in a single device.

In conclusion, we have revealed strong spin-orbit physics in ultraclean carbon
nanotube from cotunneling spectroscopy, leading to a rich competition with the
Kondo effect. Our results are in agreement with theoretical
predictions~\cite{galpin1,fang1}, if one takes into account renormalizations of
the spin-orbit interaction for two-hole filling. Disorder-free nanotubes may
come one day close to model systems, but our measurements indicate that more
theoretical work beyond the present state-of-the-art~\cite{wunsch1,rontani1} is
still needed towards a complete microscopic understanding.

\begin{acknowledgments}
We thank C. Balseiro, V. Bouchiat, P. Cornaglia, S. De Franceschi, M.-B. Lepetit, M. Rontani, 
and G. Usaj for valuable discussions, E. Eyraud, R. Haettel, D. Lepoittevin for 
technical support, J. L. Tomassin, H. Haas, T. Meunier for help in the device 
fabrication, and C. Thirion, E. Bonet, R. Piquerel for the data acquisition software
development. Devices were fabricated in the PTA (CEA/CNRS) facility. This work
is financially supported by ANR-PNANO project Mol-NanoSpin No ANR-08-NANO-002,
and ERC Advanced Grant MolNanoSpin No 226558.
\end{acknowledgments}

\end{document}